\documentclass[review]{elsarticle}

\usepackage{hyperref}
\usepackage{bm}
\usepackage{color, soul}
\usepackage{amsmath}

\journal{Computers \& Mathematics with Applications}

\begin{document}

\begin{frontmatter}

\title{An Adaptive Volumetric Flux Boundary Condition for Lattice Boltzmann Methods}

\author{James E. McClure \cortext[mycorrespondingauthor]{Corresponding author}\fnref{addr1}}
\fntext[addr1]{Advanced Research Computing, Virginia Tech, Blacksburg, USA.}
\ead{mccurej@vt.edu}

\author{Zhe Li\fnref{addr2}}
\fntext[addr2]{Department of Applied Mathematics, Australia National University, Canberra, ACT 2601, Australia}


\author{Adrian P. Sheppard\fnref{addr2}}

\author{Cass T. Miller\fnref{addr3}}
\fntext[addr3]{University of North Carolina at Chapel Hill}

\begin{abstract}
This paper presents a spatially and temporally adaptive boundary condition to specify the volumetric
flux for lattice Boltzmann methods. The approach differs from standard velocity boundary
conditions because it allows the velocity to vary over the boundary region provided
that the total flux through the boundary satisfies a prescribed constraint, 
which is a typical scenario for laboratory experimental studies.
This condition allows the boundary pressure to adjust dynamically to yield a specified boundary flow rate as a means to avoid unphysical mismatch between the boundary velocity and the interior flow field
that can arise when a constant velocity boundary condition is applied. 
The method is validated for simulation of one- and two-fluid flow in complex materials, 
with conditions determined to match typical experiments used to study flow
in porous media. 

\end{abstract}


\end{frontmatter}


\section{Introduction}

It is often desirable to design computational protocols that match particular experimental 
conditions. Setting appropriate boundary conditions is an important aspect of 
this endeavor. In computational methods, artificial boundary conditions are routinely imposed as a way to focus computational effort on a particular region of interest \cite{Colonius_04}.
Lattice Boltzmann methods (LBMs) are a broad class of computational methods that 
are used widely to study complex fluid flows \cite{Leclaire_Reggio_etal_13,Kang_Chen_etal_14,Chen_Kang_etal_14,Liu_Valocchi_etal_13,Ramstad_Idowu_etal_12,Porter_Coon_etal_12,Ahrenholz_Tolke_etal_08}. 
Boundary conditions for the LBM differ from
standard Neumann and Dirichlet boundary conditions used for partial differential equations (PDEs) because of 
the way that LBMs are constructed. The LBM originates as a discrete form of the Boltzmann
equation, and the number of unknown quantities at the boundary is determined by this choice. Boundary conditions must determine each
unknown distribution, with the total number of unknowns determined 
by the discrete velocity structure and boundary shape.
Commonly used boundary conditions for LBMs include pressure, velocity, periodic and 
outflow boundary conditions \cite{Lou_Guo_etal_13,Maier_Bernard_etal_96}.
For experimental studies of flows in porous media, microfluidics, and other complex materials,
it is common to monitor (or to control directly) the total volumetric injection rate into the system. To be specific, we will call this common volumetric flux boundary condition a macroscale condition since it is an integrated quantity applied on the boundary. The common alternative conditions are microscale conditions because these conditions prescribe point-wise values of fluid velocities or pressures at the microscale, or lattice scale. 
Under such conditions, the microscale velocity profile at the boundary will be known only on rare
occasions. Velocity boundary conditions that are inconsistent with the interior flow present 
a particular challenge, since such conditions are a source of physical inaccuracy.






When setting velocity boundary conditions, inaccuracy can result if the  condition assigned leads to a rapid change in flow conditions near the boundary region. In particular, large
gradients in an underlying potential field may result. Since potential gradients induce flow, spurious behavior
can arise to correct artifacts in the potential field.  
Since the potential and velocity cannot be independently determined,
a velocity boundary condition can lead to direct enforcement of potential gradients along the 
boundary. When the potential is determined implicitly, flow may be inconsistent with the local potential field. Setting constant microscale potential boundary conditions (e.g. a pressure boundary condition) is simpler and often more physically reasonable. However, in this scenario,
the macroscale boundary flow rate is determined as a result of the microscale system dynamics, and cannot be prescribed using established approaches. We consider the case where the total macroscale volumetric flux through a particular boundary is specified, and seek a boundary condition consistent with this condition.

Thus, the overall goal of this work is to derive a macroscale  flux boundary condition that applies to the LBM simulation of flow through porous media that is stable and efficient. The specific objectives of this paper are (1) to formulate a general boundary condition to control the volumetric flux in lattice Boltzmann
methods; (2) to validate the numerical approach based on analytical results; and (3) to apply the method to match experimental conditions for single-fluid and two-fluid flows.

\section{Methods}

LBMs are a computationally efficient class of numerical method that are widely 
used to model flows in complex geometries. Inspired by kinetic theory, LBMs solve for the evolution of a fluid flow by considering a set of distributions $f_q$,
each associated with a discrete velocity $\bm{\xi}_q$ with $q\in \{ 0,1,\ldots Q\}$.
Subject to constraints on symmetry and Gallilean invariance, LBMs have been developed
using various different discrete velocity sets to model flows in
two (e.g. D2Q9) or three dimensions (e.g. D3Q13, D3Q15, D3Q19, D3Q27) \cite{dHumieres_Lallemand_etal_1986,Qian_dHumieres_etal_1992,Chen_Martinez_etal_1992,dHumieres_Bouzidi_etal_2001}.
In this work, we present a volumentric flux boundary condition for the popular D3Q19 model.
The same general principles can be used to derive analogous boundary conditions for other
models. In the D3Q19 model, the set of discrete velocities are
\begin{equation}
 \xi_q = \left
  \{ \begin{array}{ll}
    \{ 0,0,0\}^T & \mbox{for $q=0$} \\
    \{ \pm 1,0,0\}^T,  & \mbox{for $q=1,2$} \\
    \{ 0,\pm 1,0\}^T,   &\mbox{for $q=3,4$} \\
    \{ 0,0,\pm 1\}^T  & \mbox{for $q=5,6$} \\
    \{ \pm 1,\pm 1,0\}^T, & \mbox{for $q=7,8,9,10$} \\
    \{ \pm 1,0,\pm 1\}^T, & \mbox{for $q=11,12,13,14$} \\
    \{ 0,\pm 1,\pm 1\}^T   & \mbox{for $q=15,16,17,18$}\;. \\
\end{array} \right.
\label{eq:d3q19}
\end{equation}
The distributions evolve according to the lattice Boltzmann equation
\begin{equation}
f_q(\bm{x}_i + \bm{\xi}_q \delta t,t + \delta t) = f_q(\bm{x}_i,t) + \Omega_q(\bm{x}_i,t)\;,
\label{eq:LBE}
\end{equation}
where $\bm{x}_i$ are points on a three-dimensional lattice, $i \in \{0,1,\ldots,N\}$,
$\delta t$ is the time step, and $\Omega_q(\bm{x}_i,t)$ is a collision operator that accounts
for intermolecular collisions and other interactions (as in Boltzmann's equation). The
key physics of the method are contained in the collision operation. By constructing 
different collision operators, LBMs have been constructed to recover the 
Navier-Stokes equations \cite{Benzi_Succi_etal_1992,Chen_Doolen_1998} and model a wide range of physical processes including multiphase flow \cite{Gunstensen_Rothman_etal_1991,Shan_Chen_1993,Swift_Osborn_etal_1995,He_Chen_Zhang_1999,Lee_Liu_2010}, heat transfer \cite{Shan_1997,He_Chen_etal_1998,Lallemand_Luo_2003,Li_Mei_etal_13}, diffusion \cite{Ponce_Chen_Doolen_1993,Wolf-Gladrow_1995,Sman_Ernst_2000}, reactive transport 
\cite{Zhang_Bengough_etal_2002,Verhaeghe_Arnout_etal_2006,Kang_Lichtner_etal_2007} and others. Since the basic approaches used to set boundary conditions are similar, the boundary condition developed here can be extended to other physical contexts as well.

In this work, an adapted multi-relaxation time (MRT) LB model is implemented for single-/two-fluid flow as described in McClure \textit{et al.} \cite{McClure_Prins_etal_14}, which is based on the ``color'' model initially proposed by Gustensen \textit{et al.} \cite{Gunstensen_Rothman_etal_1991}. More details of the model can be found in {\color{blue}Appendix A}. In short, an MRT formulation for a D$d$Q$q$ lattice structure models the relaxation processes individually on a set of $q$ moments determined from the distributions, where each moment relaxes toward its equilibrium value at a unique rate specified by relaxation parameters. Following the previous work by Pan \textit{et al.} \cite{Pan_Luo_Miller_2006}, the fluid kinematic viscosity $\nu$ is related to one of the relaxation parameter $\tau$ by $\nu = c_s^2(\tau-0.5)$, where $c_s$ is the LBM speed of sound. Other relaxation parameters can be found in {\color{blue}Appendix A}. 

The interpretation of the distributions is key to constructing LBMs to model different
physical phenomena. Based on this, moments of the distributions track the behavior
of physical quantities of interest. Often the distributions are defined to determine
the evolution of the number density,
\begin{equation}
\rho = \sum_{q=0}^{Q-1} f_q\;,
\label{eq:m0}
\end{equation}
and the mass flux (momentum density),
\begin{equation}
\bm{j}=\rho_0 \bm{u}= \sum_{q=0}^{Q-1} f_q \bm{\xi}_q\;,
\label{eq:m357}
\end{equation}
where $\rho_0$ is a reference density used to obtain incompressible flow. 
This represents a typical LBM formulation, although
distributions may also be defined to track other physical quantities of interest.
In the LBM, the pressure is often directly linked to the density,
\begin{equation}
p = c_s^2 \rho\;,
\label{eq:pressure}
\end{equation}
which is an expression of the ideal gas law.  
Boundary conditions are needed to determine unknown distributions along the boundary, which in turn determine
the density $\rho$ and momentum density $\rho_0 \bm{u}$.

The most familiar context for fluid flow simulations is to set pressure and\slash or velocity boundary conditions.
The basic ideas used to set pressure or velocity boundary conditions for LBMs were first
introduced by Zou and He for the D2Q9 model \cite{Zou_He_97}. 
Along a boundary region $\Gamma$, only a subset of the distributions
will be unknown. For some $\bm{x}_i \in \Gamma$, distributions $f_q$ 
are unknown for all $q$ such that $\bm{x}_i - \bm{\xi}_q \delta t \not \in D$, where $D$ is the domain. At
the inlet, the unknown distributions are: 
$f_5,f_{11},f_{14},f_{15}$ and $f_{18}$. Three of the unknown distributions can
be determined based on Eqs. \ref{eq:m0}--\ref{eq:m357}. As a consequence of the continuity 
equation, it is not possible to set both $\rho$ and $u_z$ along the $z$ inlet or outlet. 
When setting a pressure (i.e. density) boundary condition at the $z$ inlet, a consistency 
condition establishes the associated velocity 
$u_z$ as a function of the known distributions and density
\begin{eqnarray}
u_z = \frac{\rho}{\rho_0} -  \frac1 \rho_0 \Big[f_0+f_1+f_2+f_3+f_4+f_7+f_8+f_9+ f_{10} + \nonumber \\
				2(f_6+ f_{12}+f_{13}+f_{16}+f_{17}) \Big]\;.
\label{eq:inlet}
\end{eqnarray}
The consistency condition will be used to derive an adaptive pressure boundary condition that 
satisfies a specified macroscale boundary volumentric flux. 


In this work, we seek to specify the total volumetric flux across the boundary, 
which is defined as
\begin{equation}
Q_z =\int_{\Gamma_{in}} u_z d r \;,
\label{eq:flux}
\end{equation}
where $\Gamma_{in}$ is the inlet boundary.
We note that at each microscale point on the boundary $u_z$ can be determined prior to setting the pressure boundary condition based on the consistency condition. 
Combining Eq. \ref{eq:inlet} with Eq. \ref{eq:flux} we obtain
\begin{eqnarray}
Q_z = \int_{\Gamma_{in}} \frac{\rho}{\rho_0} -  \frac {1}{\rho_0}
\Big[f_0+f_1+f_2+f_3+f_4+f_7+f_8+f_9+ f_{10} + \nonumber \\
				2(f_6+ f_{12}+f_{13}+f_{16}+f_{17}) \Big] d r \;.
\label{eq:flux-1}
\end{eqnarray}
Our objective is to determine the value of $\rho$ that will produce a 
user-specified $Q_z$, where $\rho$ is constant over the boundary $\Gamma_{in}$.
The expression can be rearranged to solve for $\rho$ in terms of the known distributions on $\Gamma_{in}$
\begin{eqnarray}
\rho = \frac{\rho_0 Q_z}{A} + \frac 1 A \int_{\Gamma_{in}}
\Big[f_0+f_1+f_2+f_3+f_4+f_7+f_8+f_9+ f_{10} + \nonumber \\
				2(f_6+ f_{12}+f_{13}+f_{16}+f_{17}) \Big] d r\;,
\label{eq:potential}
\end{eqnarray}
where $A$ is the area of the inlet.
Integrating the consistency condition over the boundary thereby determines $\rho$.
As with other boundary conditions for the LBM, the condition must be applied
after streaming and prior to collision.
At each timestep, the boundary condition is set in two steps; first $\rho$ is
determined by integrating the consistency condition according to Eq. \ref{eq:potential}, then 
a pressure boundary condition is enforced in the usual way based on 
Eqs. \ref{eq:pressure} and \ref{eq:potential}.
For the pressure boundary condition, the strategy to determine the remaining two unknowns for the 
D3Q19 model is based on the work due to Hecht and Harting \cite{Hecht_Harting_10} (see more details in Appendix B). 
An analogous calculation can be performed at the outlet boundary, although
it is not necessary or advantageous to set a flux boundary condition at both boundaries.
Since the potential field is in general only known up to a constant, it 
is convenient to set a flux boundary condition at one end of the sample and rely on a pressure 
boundary condition at the other end of the domain, where the other four boundaries can be assigned using periodic or no flow conditions. We have constrained our case to match typical experimental conditions, but the notions can be extended to other sorts of systems as well.  

\section{Results}

\subsection{Single-phase Poiseuille flow}

\begin{figure}[h!]
\centering
\includegraphics[width=1.0\textwidth]{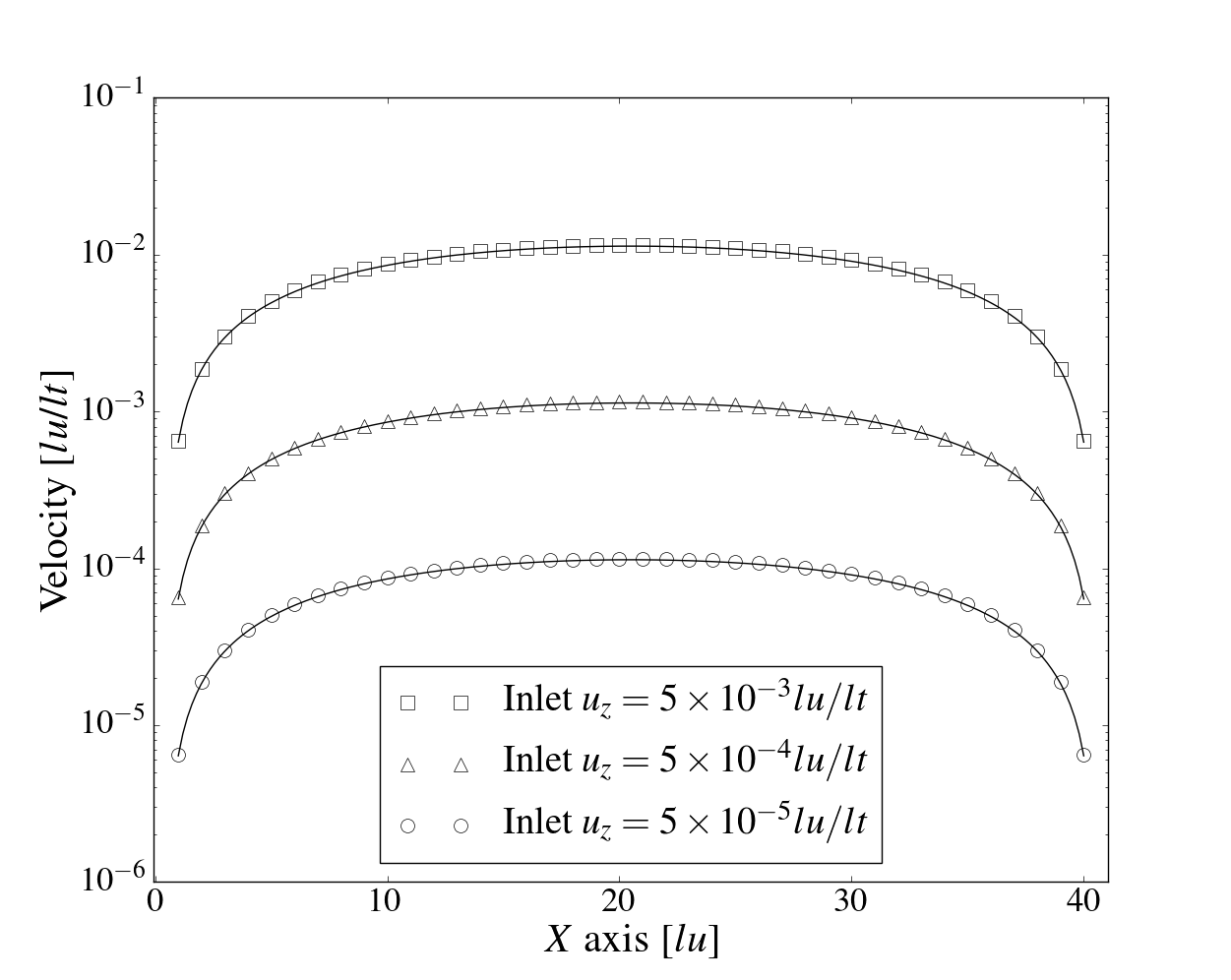}
\caption{ The velocity profiles for Poiseuille flow along the central line ($y=20$ lu) at the middle plane of the square tube ($z=40$ lu). The solid lines indicate the theoretical solutions given by Eq.\ref{eq:poiseuille_theory}. 
}
\label{fig:single_phase_poisueille_flow}
\end{figure}

In this section, to verify the accuracy of the proposed macroscale volumetric flux boundary condition, a three-dimensional (3D) Poiseuille flow simulation in a square tube of size ($L_x$, $L_y$, $L_z$) = (40, 40, 80) is performed. The discrete unit of the computational grid is a lattice unit (lu), and the iteration of the LB simulation is in the unit of lattice time (lt). The flow direction is along the $z$-axis. For a square tube, if the Cartesian origin is at the center of the plane normal to the flow axis, and the flow region is: $-w \leq x \leq w$ and $-w \leq y \leq w$, the 3D Poiseuille flow is known to have a steady-state solution given by \cite{Zhang_Shi_etal_15}:
\begin{equation}
u_z(x,y) = \frac{16 a^2}{\nu \pi^3} \left( -\frac{dp}{dz} \right) \sum_{k=1,3,5,...}^{\infty} (-1)^{(k-1)/2} \left\lbrace 1- \frac{\cosh [k \pi x/(2w)]}{\cosh (k \pi /2)} \right\rbrace \frac{\cos [k \pi y/(2w)]}{k^3}\;, 
\label{eq:poiseuille_theory}
\end{equation}
where $w$ is half of the width of the square tube, $dp/dz$ is the pressure gradient along the flow axis of the tube, and $\nu$ is the kinematic viscosity of the fluid. The infinite series in Eq.\ref{eq:poiseuille_theory} was truncated at $k=200$ to allow for a good approximation of the theoretical values. For the numerical simulations, the criterion used to determine steady state flow is
\begin{equation}
\frac{\sum_{\bm{x}} |\bm{u}(\bm{x},t)-\bm{u}(\bm{x},t-1000)|}{\sum_{\bm{x}}| \bm{u} (\bm{x},t) |} \leq 10^{-6}\;.
\label{eq:stopping_criterion_numerical_poiseuille_flow}
\end{equation}

The proposed boundary condition was incorporated into the MRT LBM with the relaxation time chosen as $\tau=1.0$. Three cases of inlet fluid velocity $u_z$ were applied to the boundary, and Fig. \ref{fig:single_phase_poisueille_flow} shows that the simulation results compared to the analytical solution given in Eq. \ref{eq:poiseuille_theory}. The numerical results are in close agreement with the analytical solutions, which validates the implementation for single-fluid flow. 

\subsection{Immiscible displacement at constant capillary number}

The LBM is often used to simulate immiscible two-fluid displacement in porous media. We consider a typical experiment in which the  following quantities are known:
\begin{enumerate}
\item $Q_z$ the volumetric flow rate (e.g. in mL/min),
\item $L_x \times L_y \times L_z$ the physical dimensions of the sample (e.g. in mm),
\item $\epsilon$ the porosity of the sample,
\item $\mu_w, \mu_n$ the dynamic viscosity for each fluid (e.g. in $\mbox{mPa} \cdot \mbox{s}$), and
\item $\gamma_{wn}$ the interfacial tension between fluids (e.g. mN/m).
\end{enumerate}
To match experimental conditions with a simulation, physical quantities must be expressed in terms of 
the lattice length $\delta x$ and the timestep $\delta t$. When the input geometry 
is provided from experimental micro-computed tomography ($\mu$CT), the lattice spacing
$\delta x$ is determined based on the width of a voxel (i.e. the image resolution). The relationship for time is obtained by considering
appropriate non-dimensional quantities and choosing the simulation parameters such that experimental
conditions are met.
For an experiment where one fluid is displacing another and compressibility effects are negligible,
the flow rates for each fluid will satisfy
\begin{equation}\label{eq:time_rate_of_Sw}
 \frac{\partial s^w}{ \partial t} = \frac{Q_z}{\epsilon V}, 
\end{equation}
where $s^w$ is the wetting-phase saturation, and $V$ is the total volume of the system. The conversion between the lattice timestep $\delta t$
and physical units can therefore be determined based on the rate of change in saturation. 
Noting that this choice does not uniquely determine the parameters, for two-fluid flows
it is desirable to match the capillary number,
\begin{equation}
\mbox{Ca} = \frac{\mu_w Q_z}{\gamma_{wn} \epsilon  A}\;,
\end{equation}
where $A$ is the area of the inlet boundary $\Gamma_i$, and the mobility
\begin{equation}
\mbox{M} = \frac{\mu_w}{\mu_{n}}\;.
\end{equation}
An additional constraint is obtained by choosing the simulated capillary number to match
the experimental value,
\begin{equation}
Q_z^{sim} = \epsilon A^{sim} \frac{\gamma_{wn}^{sim}}{\mu_{w}^{sim}} \mbox{Ca}.
\end{equation}
To reduce the number of time steps required, it is desirable to 
choose $Q_z^{sim}$ to be as large as possible, since this will induce the largest
change in saturation per time step. At fixed $\mbox{Ca}$, this is 
accomplished when $\gamma_{sim}/\mu_{sim}$ is as large as possible. The
values of $\gamma_{sim}$  and $\mu_{sim}$ are constrained by numerical
stability and the mobility; for the color-gradient based LBM used in this work \cite{McClure_Prins_etal_14}, the stable range for fluid
parameters explored was 
$1\times 10^{-5} \le \gamma_{wn}^{sim} \le 1\times 10^{-2} $, 
$ 1/15 \le \nu_i \le 1/3$ and 
$0.01 \le \rho_i \le 1.0 $ for $i \in \{w,n\}$. As a general rule of thumb, 
LBMs tend to become unstable if flow velocity ($|\bm{u}|$) exceeds $\sim 0.1$ anywhere on the lattice. 
Combinations of parameters that create this situation can result 
numerical instability (since the LBM is an explicit method) and 
compressibility errors (since the continuum physics are only recovered 
in the limit of small Mach number) \cite{Sterling_etal_96,Junk_etal_05}.

\subsubsection{Immiscible two-fluid displacement in a square tube}
\begin{figure}[h!]
\centering
\includegraphics[width=1.1\textwidth]{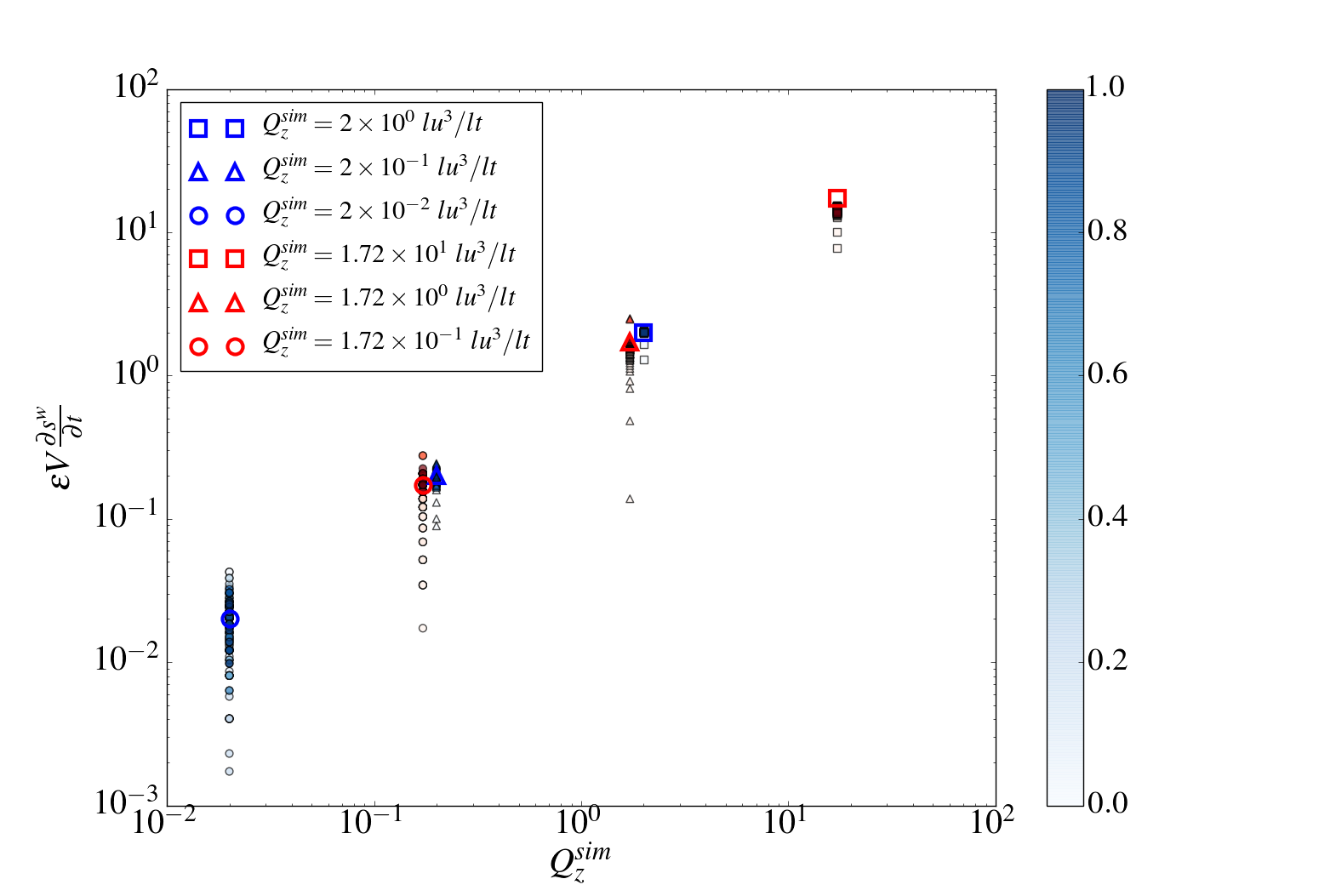}
\caption{The time rate change of the saturation (scaled by the pore volume of the media) in the primary drainage simulations, is plotted against the volumetric flow rate $Q_z^{sim}$ specified at the inlet, for the case of the square tube in blue, and for the case of the Bentheimer sandstone in red. The color of the data points indicates the temporal evolution, and the time scale is normalized by the total simulation time. For visual clarity only the color bar (in blue) for the square tube case is presented. The blank symbols are used to highlight the exact values of $Q_z^{sim}$. The lattice surface tension $\gamma_{wn}^{sim}$ is $6 \times 10^{-5}$. The phase density $\rho_i^{sim}$ is 1.0, and the phase kinematic viscosity $\nu_i^{sim}$ is $1/6$ (i.e. $\tau_i$ = 1.0), where $i \in \{w,n\}$. 
}
\label{fig:dswdt_vs_Qz_SquareTube_Bentheimer_combined}
\end{figure}

The proposed boundary condition was first investigated in a square tube where a drainage simulation was performed. The same tube size as in single-phase simulations was used. The computation domain consists of a capillary tube sandwiched by a non-wetting phase reservoir (NWR) and a wetting phase reservoir (WR), each with six layers of pure fluid nodes. For simplicity, unity density and viscosity ratios was used. Three cases of lattice volumetric flow rate, $Q_z^{sim}$ = \{0.02, 0.2, 2.0\} lu$^3$/lt, were simulated. As shown in Fig. \ref{fig:dswdt_vs_Qz_SquareTube_Bentheimer_combined}, the time rate change of the saturation, $\partial s^w / \partial t$, multiplied by the pore volume of the tube ($\epsilon V$), is plotted (in blue) against different $Q_z^{sim}$. The color of the data points indicates the temporal evolution, with the time scale normalized by the total simulation time. It can be seen that, as time evolves, the time rate change of the saturation approaches the prescribed $Q_z^{sim}$ once the steady state displacement is reached. It is also noted that at the initial stage $\partial s^w / \partial t$ deviates from $Q_z^{sim}$, but eventually stabilizes to match the to match the boundary flux. While the prescribed boundary flux will match exactly (since Eq. \ref{eq:potential}
is not approximate),
fluctuations in $\partial s^w / \partial t$ are possible due to fluid compressibility and the rearrangement of the diffuse interface in the color LBM. 
At low flow rates
the presence of spurious currents may influence the accuracy of the boundary condition, which 
is a known limitation of the color LBM. This can be mitigated by using
larger fluid reservoirs such that spurious currents do not arise in proximity to the boundary. 

\subsubsection{Immiscible two-fluid displacement in a realistic porous medium}

\begin{figure}[h!]
\includegraphics[width=0.8\textwidth]{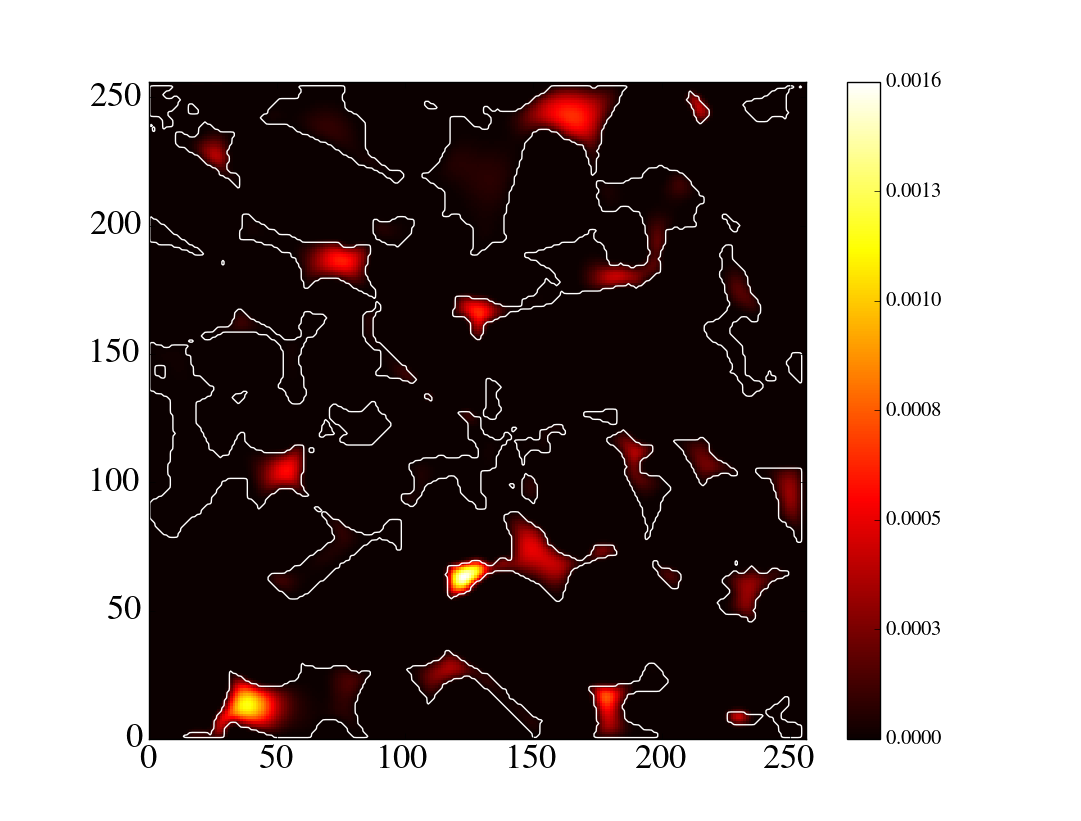}
\caption{The cross-sectional view of the velocity field $u_z(x,y)$ at the inlet boundary of NWR in Bentheimer sandstone primary drainage simulation for the case of $Q_z^{sim}$ = 1.72 lu$^3$/lt. The white contour line depicts the fluid-solid boundary of the first layer of the medium. The velocity field was extracted at time step 250,000 lt when the steady state displacement was reached.
}
\label{fig:bentheimer_2d_vel_heat_map}
\end{figure}

The proposed boundary condition was also tested with primary drainage simulations in an X-ray $\mu$CT image of Bentheimer sandstone sample. A sub-domain of $256^3$ lu$^3$ of the original image was used, with an image resolution of 4.95 $\mu$m/lu \cite{Herring_Middleton_etal_17}. The sub-domain was again sandwiched by six layers of NWR and WR, respectively. Unity density and viscosity ratios were adopted. Three cases of lattice volumetric flow rate, $Q_z^{sim}$ = \{0.172, 1.72, 17.2 \} lu$^3$/lt were set such that the capillary numbers were the same as in the square tube case. The corresponding time rate change of the saturation is also shown in Fig. \ref{fig:dswdt_vs_Qz_SquareTube_Bentheimer_combined} in red. Due to the initial capillary entry effect, the time rate change gradually approaches the prescribed $Q_z^{sim}$ as the steady state displacement is reached. Moreover, to illustrate the capability of the proposed boundary condition to locally adjust the inlet flux, the two-dimensional $u_z(x,y)$ profile at the inlet boundary of NWR, for the case of $Q_z^{sim} = 1.72$ lu$^3$/lt is shown in Fig. \ref{fig:bentheimer_2d_vel_heat_map}. Since the NWR consists of pure fluid nodes, a contour line in white delineating the fluid-solid boundary of the first layer of the porous medium is also shown. It can be seen that the proposed boundary condition only directs positive flux towards the pore space of the medium, while maintaining zero flux for where the solid phase is present. This demonstrates that the boundary condition allows the local flow rate to vary across the boundary region based on the interior structure of the flow, while maintaining control over the volumetric
flow rate for fluid injected into the system.


\section{Conclusions}

In this paper, we present a volumetric flux boundary condition for lattice Boltzmann methods.  
The approach is derived based on a consistency condition that is associated with a pressure boundary 
condition. By integrating the consistency condition over the relevant boundary region, a spatially-constant
potential can be determined and enforced along that boundary to produce a desired volumetric flow rate.
The local velocity can vary in time and space along the boundary depending on the 
interior flow dynamics, providing an advantage relative to the standard velocity boundary conditions used in conjunction
with LBMs. The boundary condition is validated analytically for one-and two-fluid lattice Boltzmann schemes and applied to simulate two-fluid flow within an experimentally-obtained Bentheimer 
sandstone image. We delineate an approach to match experimental conditions for two-fluid flow based on the
resolution, volumetric flow rate, capillary number and viscosity ratio. 
For two-fluid simulations, spurious currents associated with the interfacial stresses can reduce the accuracy 
of the approach, although the method is sufficient to set the capillary number to match
experimental conditions in practice. The boundary condition provides an attractive alternative to existing
LBM boundary conditions for modeling flow experiments within porous media.

\section*{Appendix A: Momentum and mass transport in multiphase lattice-Boltzmann model}
\label{appendix:A}
The multiphase ``color'' LBM used in this work is based on the implementation described in McClure \textit{et al.} \cite{McClure_Prins_etal_14}. The momentum transport is modeled by the lattice-Boltzmann equation (LBE) as:
\begin{equation*}
f_q(\bm{x}_i + \bm{\xi}_q \delta t,t + \delta t) - f_q(\bm{x}_i,t) = \sum^{Q-1}_{k=0} M^{-1}_{q,k} S_{k,k} (m_k^{eq}-m_k)\;,
\end{equation*}
where the transformation matrix $M_{q,k}$ (its inverse $M^{-1}_{q,k}$) maps the distribution function to its moments by $m_k = \sum_{q=0}^{Q-1} M_{q,k} f_q$, and diagonal matrix $S_{k,k}$ specifies the relaxation rates for each moment. For D3Q19 lattice structure, the $M_{q,k}$ can be found in \cite{dHumieres_Ginzburg_etal_2002}, and the 19 moments are defined as:
\begin{equation*}
\bm{m} = (\rho,e,\epsilon,j_x,q_x,j_y,q_y,j_z,q_z,3p_{xx},3\pi_{xx},p_{ww},\pi_{ww},p_{xy},p_{yz},p_{zx},m_x,m_y,m_z)\;,
\end{equation*}
These 19 moments \{$m_k \mid k=0,1,...,18$\} are the mass density ($m_0=\rho$), the part of the kinetic energy independent of the density ($m_1=e$), the part of the kinetic energy square independent of the density and kinetic energy ($m_2=\epsilon = e^2$), the momentum flux ($m_{3,5,7}=j_{x,y,z}$), the energy flux ($m_{4,6,8}=q_{x,y,z}$), the symmetric traceless viscous stress tensor ($m_9=3p_{xx}$, $m_{11}=p_{ww}$, and $m_{13,14,15}=p_{xy,yz,zx}$), the vectors of quartic order ($m_{10}=3 \pi_{xx}$, $m_{12} = \pi_{ww}$), and the vectors of cubic order ($m_{16,17,18}=m_{x,y,z}$) \cite{dHumieres_Ginzburg_etal_2002}. The relaxation rates for each moment are given by:
\begin{equation*}
\bm{S}=\text{diag}(0,s_e,s_{\epsilon},0,s_q,0,s_q,0,s_q,s_{\nu},s_{\pi},s_{\nu},s_{\pi},s_{\nu},s_{\nu},s_{\nu},s_m,s_m,s_m)\;,
\end{equation*}
where, the relaxation rates for the conserved moments, the density $\rho$ and the momentum $(j_x,j_y,j_z)$, are set to zero, since they are not affected by collisions. Following the reported work in \cite{dHumieres_Ginzburg_2003}, the relaxation rates for the non-conserved moments are set as
\begin{equation*}
s_e=s_{\epsilon}=s_{\pi}=s_{\nu},\,\,\,s_q=s_m=8\frac{(2-s_{\nu})}{(8-s_{\nu})}\;.
\end{equation*}

The fluid kinetic viscosity $\nu$ is given by:
\begin{equation*}
\nu = c_s^2 (\frac{1}{s_{\nu}}-\frac{1}{2})\;,
\end{equation*}
and in the main text, the commonly used relaxation time $\tau$ is defined as $\tau = s_{\nu}^{-1}$. 

In the case of multiphase flow, the equilibrium moments $m_q^{eq}$ are set such that the stress tensor matches that of a Newtonian fluid with an anisotropic contribution due to the interfacial tension. Following McClure \textit{et al.}, the non-zero equilibrium moments are given by: \cite{McClure_Prins_etal_14}
\begin{eqnarray*}
m_1^{eq} &=& (j_x^2+j_y^2+j_z^2) + \alpha |\textbf{C}| \\
m_9^{eq} &=& (2j_x^2-j_y^2-j_z^2)+ \alpha \frac{|\textbf{C}|}{2}(2n_x^2-n_y^2-n_z^2) \\
m_{11}^{eq} &=& (j_y^2-j_z^2) + \alpha \frac{|\textbf{C}|}{2}(n_y^2-n_z^2) \\
m_{13}^{eq} &=& j_x j_y + \alpha \frac{|\textbf{C}|}{2} n_x n_y \\
m_{14}^{eq} &=& j_y j_z + \alpha \frac{|\textbf{C}|}{2} n_y n_z \\
m_{15}^{eq} &=& j_x j_z + \alpha \frac{|\textbf{C}|}{2} n_x n_z\;, \\
\end{eqnarray*}
where the parameter $\alpha$ is linearly related to the interfacial tension, and $\textbf{C}$ is the color gradient, which is defined as the gradient of the phase field:
\begin{equation*}
\textbf{C}=\nabla \varphi\;,
\end{equation*}
where the phase field $\varphi$ is defined based on the densities of the non-wetting and wetting fluids, $\rho_n$ and $\rho_w$, respectively, which is given by:
\begin{equation*}
\varphi = \frac{\rho_n-\rho_w}{\rho_n+\rho_w}\;.
\end{equation*}
$\bm{n} = (n_x,n_y,n_z)$ is the unit normal vector of the color gradient and is calculated as:
\begin{equation*}
\bm{n} = \frac{\textbf{C}}{|\textbf{C}|}\;.
\end{equation*}
The phase indicator field is tracked by solving two additional mass 
transport LBEs that rely on the three-dimensional, seven velocity 
model (D3Q7). The seven velocities for the D3Q7 model correspond to 
$q=0,1,\ldots,6$ in the D3Q19 model. D3Q7 distributions 
model the evolution of the number density of each fluid, $N_A$ and $N_B$, respectively,
which are given by
\begin{equation}
N_A = \sum_{q=0}^6 A_q\;, \qquad
N_B = \sum_{q=0}^6 B_q\;, \hbox{ and }
\phi = \frac{N_A - N_B}{N_A+N_B}\;.
\end{equation}
The distributions are updated based on
\begin{eqnarray}
A_q(\bm{x} + \bm{\xi}_q \delta t, t+\delta t) &=& w_q N_A \Big[1 + \frac 92 \bm{u} \cdot \bm{\xi}_q 
+ \beta  N_B \bm{n} \cdot \bm{\xi}_q\Big] \hbox{ and} 
\\
B_q(\bm{x} + \bm{\xi}_q \delta t, t+\delta t) &=& 
w_q N_B \Big[1 + \frac 92 \bm{u} \cdot \bm{\xi}_q 
- \beta  N_A \bm{n} \cdot \bm{\xi}_q\Big]\;, 
\end{eqnarray}
where $\beta$ controls the interface width, $w_0=1/3$ and $w_{1,\ldots,6} = 1/9$. 
The mass transport LBEs ensure phase separation based on the color gradient,
which then couples to the momentum transport. 

\section*{Appendix B: Pressure Boundary Condition for D3Q19}
\label{appendix:B}
At the inlet, the unknown distributions are $f_5, f_{11}, f_{14}, f_{15}$ and 
$f_{18}$. The above expressions can be rearranged to place the unknowns on 
the left-hand side:
\begin{eqnarray*}
f_{5} + f_{11}+ f_{14} + f_{15}+ f_{18} &=& \rho- (f_{0} + f_{1} + f_{2} + f_{3} + f_{4}  + f_{6} 
			+ f_{7} + \nonumber \\ && f_{8} + f_{9} + f_{10} 
            + f_{12} + f_{13} 
             + f_{16} + f_{17} )\\
 f_{11}- f_{14}  &=& \rho_0 u_x- (f_{1} - f_{2}
			+ f_{7} - f_{8} + f_{9} - f_{10} 
             - f_{12} + f_{13} )\\
 f_{15}- f_{18} &=& \rho_0 u_y -(f_{3} - f_{4} 
			+ f_{7} - f_{8} - f_{9} + f_{10}
            - f_{16} + f_{17}  )\\      
 f_{5}+ f_{11}+ f_{14}+ f_{15}+ f_{18} &=&\rho_0 u_z + ( f_{6} 
             +f_{12} + f_{13} + f_{16} + f_{17} )\;. 
\end{eqnarray*}
It is clear that the sum $f_{5}+ f_{11}+ f_{14}+ f_{15}+ f_{18}$  
is determined either by choosing $\rho$ or by choosing $\rho u_z$; 
both conditions cannot be set independently. If a pressure boundary 
condition is use to determine $\rho$, then a consistency condition can 
be established by eliminating the sum of the unknowns from 
\begin{eqnarray}
\rho & - (f_{0} + f_{1} + f_{2} + f_{3} + f_{4}  + f_{6}
+ f_{7} + f_{8} + f_{9} + f_{10} + f_{12} + f_{13} + f_{16} + f_{17}) =
\nonumber \\ 
& \rho_0 u_z - ( - f_{6} - f_{12} - f_{13} - f_{16} - f_{17} )\;,
\end{eqnarray}
which can then be solved to determine the associated velocity
\begin{equation}
 u_z  = \frac \rho{\rho_0} - \frac 1\rho_0 [f_{0} + f_{1} + f_{2} + f_{3} + f_{4} 
+ f_{7} + f_{8} + f_{9} + f_{10} 
          + 2( f_{6} +  f_{12} +f_{13} + f_{16}+ f_{17} ) ]\;.
\end{equation}

The equilibrium distributions for the D3Q19 model are
\begin{equation}
f_q^{eq} (\rho,  \bm{u}) = w_i \Big[ \rho + \rho_0 \Big( 3 \bm{\xi}_q \cdot  \bm{u}
+ \frac 92 (\bm{\xi}_q \cdot \bm{u})^2 +\frac 32  \bm{u} \cdot  \bm{u} \Big)
\Big]\;.
\end{equation}
With both $\rho$ and $\mathbf{u}$ known, the unknown distributions are chosen 
by assuming that the bounce-back rule applies
to the non-equilibrium part of the unknown distributions, for example:
\begin{equation}
f_q - f_q^{eq} = f_{\overline{q}} - f_{\overline{q}}^{eq}\;, 
\end{equation}
where $\xi_q = - \xi_{\overline{q}}$.
This can be solved for the unknown distribution
\begin{eqnarray}
f_q &=& f_{\overline{q}} + f_q^{eq} - f_{\overline{q}}^{eq} \\
 &=& f_{\overline{q}} + 6 \rho_0 w_i (\bm{\xi}_q \cdot \bm u)\;,
\end{eqnarray}
where the definition of the equilibrium distributions has been inserted,
using the fact that $\xi_q = - \xi_{\overline{q}}$. This is used to 
determine
\[
f_5 = f_6 + \frac 1 3 \rho_0 u_z\;.
\]
This leaves four remaining unknowns and only three equations. Hecht and Harding 
resolve the closure problem by defining
\begin{eqnarray}
N_x^z &=& \frac 12 [f_1 + f_7 + f_9 - (f_2+f_{10}+f_{8})] - \frac 1 3 \rho_0 u_x \\
N_y^z &=& \frac 12 [f_3 + f_7 + f_{10} - (f_4+f_{9}+f_{8})] - \frac 1 3 \rho_0 u_y\;,
\end{eqnarray}
and then providing a closed system based on the equations
\begin{eqnarray}
f_{11} - f_{11}^{eq} &=& f_{12}  - f_{12}^{eq} - N_x^z\\
f_{14} - f_{14}^{eq} &=& f_{13}  - f_{13}^{eq} - N_x^z\\
f_{15} - f_{15}^{eq} &=& f_{16}  - f_{16}^{eq} - N_x^z\\
f_{18} - f_{18}^{eq} &=& f_{17}  - f_{17}^{eq} - N_x^z\;,
\end{eqnarray}
which can be simplified to the form
\begin{eqnarray}
f_{11} - f_{12}  &=&  \frac 16 \rho_0 (u_x + u_z) - N_x^z\\
f_{14} - f_{13}  &=&  \frac 16 \rho_0 (-u_x + u_z) + N_x^z\\
f_{15} - f_{16}  &=&  \frac 16 \rho_0 (u_y + u_z) - N_y^z\\
f_{18} - f_{17}  &=&  \frac 16 \rho_0 (-u_y + u_z) + N_y^z\;.
\end{eqnarray}
These expressions can be rearranged to solve for the unknown distributions
for either the inlet or outlet.

\section*{Acknowledgements}
This work was supported by Army Research Office grant W911NF-14-1-02877 and National Science Foundation grant 1619767. An award of computer time was provided by the Department of Energy INCITE program. This research also used resources of the Oak Ridge Leadership Computing Facility, which is a DOE Office of Science User Facility supported under Contract DE-AC05-00OR22725. Z.L. acknowledges the Australian Government Research Training Program (RTP) Scholarship and the Robert and Helen Crompton travel fund. A.P.S. acknowledges the support of an Australian Research Council Future Fellowship through project FT100100470. This research also used resources and services from the National Computational Infrastructure (NCI), which is supported by the Australian Government.

\section*{References}

\bibliography{BCs}

\end{document}